*Full Length Research*

# Statistics on Open Access Books Available through the Directory of Open Access Books

Dr. Keita Tsuji

Associate Professor, Faculty of Library, Information and Media Science, University of Tsukuba, Address: 1-2 Kasuga, Tsukuba-city, Ibaraki-ken 305-0821, Japan. E-mail:keita@slis.tsukuba.ac.jp

Open Access (OA) books available through the Directory of Open Access Books (DOAB) are investigated and the number of titles, the distribution of subjects, languages, publishers, publication years, licensing patterns, etc., are clarified. Their chronological changes are also shown. The sample comprised 10,866 OA books, which were available through the DOAB as of February 24, 2018. The results show that OA books are increasing in number at an accelerating rate. As for distribution of subjects, Social Sciences ("H" in the Library of Congress Classification [LCC] codes), Science ("Q" in LCC) and World History and History of Europe, Asia, Africa, Australia, New Zealand, etc. ("D" in LCC) are the most popular. As for languages, English, French, and German are the most popular. As for publishers, Frontiers Media SA, Presses universitaires de Rennes, and ANU Press are the most popular. Many books are newly published ones, but older books, published in or before 1999, also began to be available recently. As for the licensing patterns, "CC by-nc-nd" and "CC by" are the most popular. Considering these tendencies, libraries should begin to utilize OA books.

## INTRODUCTION

According to the International Federation of Library Associations (IFLA), "Open access is the now known name for a concept, a movement and a business model whose goal is to provide free access and re-use of scientific knowledge in the form of research articles, monographs, data and related materials" (IFLA, 2011). Open access (henceforth OA) has already been popular in the field of journal papers, and now it has started to gain a certain position in the field of monographs, or books. In the present paper, OA books are investigated, and the number of titles and the distribution of subjects, languages, publishers, publication years, licensing patterns, etc., will be clarified. Their chronological changes will also be shown.

Although the definition of OA books and OA monographs has not been clearly stated,[1] many articles and reports have been published on them. In the present paper, we define OA books as books of research output that are in electronic form and available on the Web free of charge.

---

[1] For instance, Adema (2012) stated that, prior to the user needs analysis of OA books, online discussion of OA books took place among publishers, academics, librarians, and participants from the wider OA and publishing community. The discussion included the definition of OA books, but Adema did not report any conclusion regarding this definition.

There are two representative directories or repositories of OA books. One is the Directory of Open Access Books (DOAB), and the other is the Open Access Publishing in European Networks (OAPEN) Library. Both are provided by the OAPEN Foundation. The DOAB is a discovery service for peer-reviewed OA books, with links to the full texts of the publications at the OAPEN Library, publisher's website, or repositories (Adema, 2012; Karak & Mandal, 2017). The DOAB was officially launched on July 1, 2013, at the Open Access Monographs in the Humanities and Social Sciences Conference at the British Library in London (Ferwerda. 2014; Karak & Mandal, 2017). However, it had been operating unofficially since 2011, and its beta version was launched in 2012. Lamani (2018) stated, "The Directory of Open Access Books (DOAB) is the major milestone in facilitating organized access to open access e-books."

As of February 24, 2018, the DOAB was providing 10,866 titles, and as of May 11, 2018, the OAPEN Library was providing 5,048 titles. In the present paper, we investigate all 10,866 titles provided by the DOAB. We leave it for future research to investigate OA books provided by the OAPEN Library. One reason is that the number of titles provided by the OAPEN Library is smaller than that provided by the DOAB. Furthermore, OA books available through the DOAB have a higher degree of *Open Access*; i.e., they have higher reusability than those in the OAPEN Library in the licensing sense. Snijder (2013) stated that Amsterdam University Press had placed almost 400 books in the OAPEN Library and that less than half of them were also available through the DOAB. He said that only books with a license that enabled readers to share the contents were allowed in the DOAB. Although it is an interesting direction to compare titles available through the DOAB and the OAPEN Library or to investigate the disjunction of OA books available through them, we leave it for future research.

There are three papers that have investigated the OA books available through the DOAB (no paper was found on those available through the OAPEN Library). Karak and Mandal (2017) investigated OA books in the field of Library and Information Science. They found 35 books and analyzed (1) the year-wise distribution and growth, (2) the number of authors, (3) publishers who were actively involved in the DOAB, (4) language-wise distribution, (5) licensing patterns, and (6) the number of pages. Khanchandani and Kumar (2017) investigated 1,052 OA books available through the DOAB in the field of Science and Technology. They analyzed the books' publishers, licenses, and language-wise distribution. Lamani et al. (2018) investigated 1,200 OA books available through the DOAB in the field of Social Sciences. They analyzed the books' subjects, the number of authors, the licenses, languages, publication years, and pagination-wise distribution. However, few studies have been done on all the titles available through the DOAB regardless of their field. As we mentioned above, we will investigate all 10,866 titles available through the DOAB. The results clarify the current status of OA books and can be used for discussion of the preferable future of OA books.

## OBJECTIVES OF THE STUDY

The main objectives of the present study are:

➢ To show the statistics on OA books available through the DOAB such as their distribution of the subjects, languages, publishers, publication/add-on years, licensing patterns.
➢ To find out the chronological changes of the above mentioned data for discussion of the preferable future of OA books.
➢ To identify the active OA books publishers to promote their further contribution.

## LITERATURE REVIEW

In this section, we will outline the papers related to OA books. The preceding papers can be classified into six types: (A) those describing the current status and problems of OA books in one institution or in general (Snijder, 2013; Hacker, 2014; Crossick, 2015; Collins & Milloy, 2016; Amano, 2017; Chakrabarti & Mandal, 2017; Hacker & Corrao, 2017; Speicher, 2017; Tanabe, 2017; Mongeau, 2018); (B) those investigating the impacts of OA books on sales of printed books and on academic fields (Ferwerda et al., 2013; Snijder, 2016; Speicher, 2016; Neylon et al., 2018); (C) those studying the users or usages of OA books (Adema, 2012; Ferwerda et al., 2013; Montgomery et al., 2017); (D) those focusing on business models (Adema, 2010; Ferwerda, 2014; London Economics, 2015); (E) those emphasizing the direction that OA books should take in the future (Jisc Collections & OAPEN Foundation, 2016; Barnes et al., 2017); and (F) those investigating the bibliographic information of OA books (Karak & Mandal, 2017; Khanchandani & Kumar, 2017; Lamani et al., 2018). Since we have already outlined type (F) above, we will introduce the rest, i.e., from (A) to (E), in the following paragraphs.

Among papers of type (A), Snijder (2013) reported on the relationship between Amsterdam University Press (AUP), the OAPEN Library, the DOAB, and IMISCOE (International Migration, Integration and Social Cohesion). He stated that the OAPEN Library and DOAB platforms were very useful tools for promoting OA books. Hacker (2014) described the challenges in publishing OA books at the University of Heidelberg and pointed out the importance of collaboration between scholars and publishers. Crossick (2015) showed that the so-called monograph crisis did not exist and examined (1) the advantages and drawbacks of the various models for OA monographs, (2) the strengths of the print monograph, (3) the problems of licensing, (4) third-party rights, (5) the implications for other stakeholders (such as publishers, learned societies, universities, and university libraries), (6) international cooperation issues, and (7) suggestions for policymakers. Collins and Milloy (2016) reported the main findings from the OAPEN-UK research project, a five-year study into OA monograph publishing in the Humanities and Social Sciences. They first referred to the perspectives of five main groups (researchers, institutions, publishers, learned societies, and funders). Then they referred to the technical and organizational elements of producing an OA monograph and the characteristics that a successful business model of an OA monograph would have. Amano (2017) described the current status and challenges of OA books in the fields of Humanities and Social Sciences in Europe, referring to their various business models, dissemination, and preservation. Chakrabarti and Mandal (2017) investigated the growth, languages, copyright licensing, and publishers of 35 e-books in the Library and Information Science field available through the DOAB. Hacker and Corrao (2017) reported the challenges, accomplishments, and setbacks that Heidelberg University's newly founded OA publisher, heiUP, experienced. They discussed issues such as acquiring manuscripts, designing and building workflows, and building an outlet for the finished product. Speicher (2017) examined the current status of OA monographs and UK university presses. Tanabe (2017) examined the programs by Springer on OA and scholarly books. Mongeau (2018) described the current models, trends, and issues of OA monograph publishing.

Among papers of type (B), Ferwerda et al. (2013) reported the results of OAPEN-NL, which was a project to gain experience in the OA publication of monographs in the Netherlands. OAPEN-NL examined (1) user needs and perceptions about the OAPEN's publishing model for OA monographs, (2) the costs of monograph publishing in the Netherlands, and (3) the effects of OA on sales and scholarly impact. They stated that OAPEN-NL found no evidence of an effect of OA on sales (since Ferwerda et al. also investigated user needs and perceptions, their paper also falls into type [C]). Snijder (2016) investigated whether OA had a positive influence on the number of citations and tweets a monograph received. He found a slight OA advantage and that it depended on the language and subjects of the books. Speicher (2016) reported the effect on print sales of OAPEN-NL and OAPEN-UK/Jisc making a book OA. She also described the experience of the University College London Press and stated that OA seemed to have little effect on print sales. Neylon et al. (2018) investigated the extent to which OA books could be seen by the communities that might make use of them. They focused on OA books made available by publishers and platforms that were part of the OPERAS network, which was focused on the development of a European research infrastructure.

Among papers of type (C), Adema (2012) reported on DOAB users' needs, expectations, and experiences of OA. These covered the awareness of the importance of OA; quality control, especially, peer-reviews; licensing; business models; and the DOAB itself. Montgomery et al. (2017) investigated the usage of OA books via the JSTOR Platform. They investigated where the readers came from, the most popular subjects of the books, readers' behavior when they downloaded or viewed books, and publishers' perspectives.

Among papers of type (D), Adema (2010) investigated various initiatives' OA business models, publishing models, and publishing processes. Initiatives included commercial publishers and presses established by societies, academies, libraries, and universities, etc. Ferwerda (2014) gave an overview of six business models of OA books: (1) hybrid models (which provided free versions of publications and the sale of premium editions), (2) institutional support (the receipt of direct financial support through grants or indirect support through subsidies from the parent institute, etc.), (3) author-side publication charges (derived from the Article Processing Chargemodel for OA journals), (4) library-side models (which used library budgets to support OA publications), (5) crowdfunding, and (6) green OA. London Economics (2015) examined business models for OA monographs and assessed issues relating to cost recovery, quality control, and the incentives for authors and publishers. They identified six types of business model, such as those by traditional publishers and new university presses.

Among papers of type (E), Jisc Collections and the OAPEN Foundation (2016) reported on the potential centralized services that would support and encourage the publication of OA peer-reviewed monographs and presented recommendations for their establishment. Barnes et al. (2017) conducted a survey of academic libraries in the United States and concluded that library-funded OA book initiatives could successfully scale up if they employed sustainable business models, offered quality content, and provided participants with usage data.

## METHODOLOGY

In this section, we will explain the data and aspects we investigated concerning DOAB OA books.

### Data We Used

The DOAB said that it had "10866 academic peer-reviewed books and chapters from 256 publishers," as of February 24, 2018.[2] We downloaded a CSV file from DOAB on February 25, 2018.[3] The file contained the works' "Title," "ISBN," "Authors," "Pages," "Publisher," "Languages," and "Year of publication," etc. It also contained information about (1) dates on which the books or chapters were added to the DOAB, as "Added-on date," and (2) the Library of Congress Classification (henceforth "LCC") codes as "Subjects."

---

[2] https://www.doabooks.org/doab?func=search&uiLanguage=en

Books and chapters are not distinguished in the above-mentioned CSV file. If we check each item manually, we may be able to distinguish books from chapters, but doing so would be quite labor-intensive. Here, we would like to recall UNESCO's definition of a book: "a non-periodical printed publication of at least 49 pages" (UNESCO, 1964). Among the 10,866 items (i.e., books or chapters) in the CSV file, (1) for "Pages," 4,567 items were not available (i.e., their data were NULL), and (2) among the rest, 6,299, only 126 (i.e., 2.0%) were less than 49 pages in length. We regard 2.0% as sufficiently small, and the results concerning all 10,866 items provide a good image or approximation of OA books. We did not remove the 126 chapters from the 6,299 items because doing so would require us to remove chapters from the other 4,567 items for consistency. That would be very labor-intensive, and we think that the effect on our results by such a removal would be small. Henceforth we will refer to both books and chapters as "titles" and show the results for them. We regard the results for titles to be very close to the results for books, as we previously mentioned.

**Aspects We Investigated**

First, we investigated the 10,866 titles' subjects, languages, publishers, publication years, licensing patterns, and years when the titles were added on the DOAB (henceforth, the "added-on years"). These data were obtained from the columns "Subjects," "Languages," "Publisher," "Year of publication," "License," and "Added-on date," respectively. We also investigated any chronological changes and the distribution of combinations of subjects, languages, and publishers. Furthermore, we randomly selected 40 English titles and investigated (a) the affiliations and positions of the first authors/editors of the titles and (b) the availability of the contents as text format.

As for the subjects, we only used the first characters of the LCC subjects. For instance, if a book was assigned the LCC subject "JF20-2112," we regarded the subject of the book to be just "J" (Political Science). When multiple LCC subjects were assigned to one book, we adopted the first one as its LCC subject. For instance, if a book was assigned LCC subjects such as "HB1-3840; D1-2009," we regarded the subject of the book to be "H" (Social Sciences). The representations of the language of each title in the DOAB CSV file were somewhat noisy. We regarded the character sequences in the right-hand column in Table 1 to represent the language shown in the left-hand column.

**RESULTS**

In this section, we will show the results for (1) the number of titles, (2) distribution of subjects, (3) distribution of languages, (4) combination of subjects and languages, (5) combination of publication and added-on years, (6) distribution of publishers, (7) licensing patterns, (8) authors and editors, and (9) availability of contents as text format, in that order.

**Number of Titles**

As we previously mentioned, the number of titles available through the DOAB as of February 24, 2018, was 10,866. On the basis of the added-on years, the numbers of titles available through the DOAB as of 2011, 2012, 2013, 2014, 2015, 2016, and 2017 were estimated to be 485, 1,173, 1,529, 2,425, 3,719, 5,698, and 10,310, respectively. Figure 1 shows these numbers. We can see in Figure 1 that the number of titles available through DOAB increases at an accelerating rate.

**Distribution of Subjects**

The subject-wise numbers of titles available through the DOAB from 2011 to 2017 and those as of February 24, 2018, are shown in Table 2. In Table 2, the leftmost "A" to "Z" represent the LCC subjects.[4] The subjects that the letters represent are shown in Table 3. For instance, Table 2 and Table 3 show that the number of titles in Political Science ("J") available through the DOAB as of 2013 is 181.

We can see in Table 2 that the most popular subject as of February 24, 2018 is Social Sciences ("H"), which accounts for 8.8% (952 titles) of OA books available through the DOAB. The second-most and the third-most popular subjects as of February 24, 2018, are Science ("Q") and World History and History of Europe, Asia, Africa, Australia, New Zealand, etc. ("D"), respectively.

The numbers of titles of subjects "A" to "K" and "L" to "Z" for 2011 to 2017 are shown in Figures 2 and 3, respectively. We can see in these figures that the number of titles in Social Sciences ("H") was the largest most of the time. The numbers of titles in Education ("L") and Medicine ("R") have increased sharply in recent years.

**Distribution of Languages**

The numbers and percentages of titles of each language available through the DOAB as of February 24, 2018 are shown in Table 4. We can see in Table 4 that the most popular language was English, which accounts for 49.0% (5,416 titles) of titles available through the DOAB. French, German, Portuguese, and Spanish follow. These five languages account for 92.2% of titles available through the DOAB.[5,6]

---

[3] https://www.doabooks.org/doab?func=about&uiLanguage=en

[4] https://www.loc.gov/catdir/cpso/lcco/

Table 1: Languages and Their Corresponding Character Sequences

| Language | Character Sequences |
|---|---|
| English | en, eng, Eng, english, englisch (eng), Englilsh, Englisch, Inglese, ENGLISH |
| French | fr, fra, fre, Fr, Fre, französisch (fre), Francais, Francese |
| German | de, german, Deutsch, Deutsche, deutsch (ger) |
| Italian | it, ita, italian, italiano, Italiano, ITALIANO |
| Spanish | es, spa, ES, Espanol, Spagnolo |
| Portuguese | pt, por, Pt, PT, Portugese |
| Slovene | sl, slv, Slovenian |
| Latin | la, lat |
| Turkish | tr |
| Romansh | rm |
| Russian | ru |
| Arabic | ar |
| Greek | el |
| Bulgarian | bg |
| Church Slavic | cu |
| Catalan | ca |

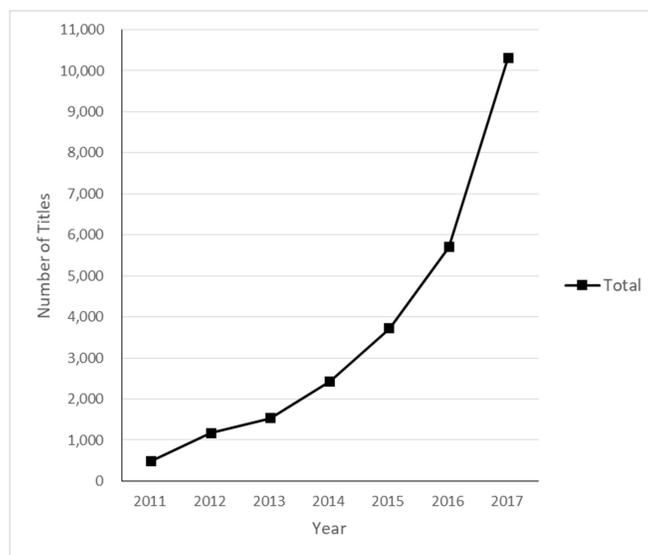

Figure 1: Changes in the Number of Titles Available through the DOAB

The numbers of titles in English, French, German, Portuguese, Spanish, Italian, Dutch, and the rest ("Other") available through the DOAB during the period from 2011 to 2017 and those as of February 24, 2018, are shown in Table 5. We can see in Table 5 that, for instance, the number of titles written in French as of 2014 was 158. Figure 4 was generated on the basis of the numbers from Table 5. We can see in Figure 4 that the number of titles written in English was consistently the largest throughout the period. The number of titles written in French increased sharply during from 2016 to 2017 and overtook those in German and Portuguese.

---

[5] Some figures in Table 4 correspond to the same titles. For instance, if a particular book has been published in English, French, and German, these editions were counted as three independent books written in English, French, and German. This is why the "Total" in Table 4 is 11,058, which is larger than 10,866 in Table 2.

[6] The DOAB CSV file said that two titles were literally written in an "Other" language. This "Other" is different from "Other" in Tables 4, 5, and 6 and Figure 4.

Table 2: Subject-wise Numbers of Titles

|   | 2011 | 2012 | 2013 | 2014 | 2015 | 2016 | 2017 | Feb. 2018 |   |   |   |
|---|---|---|---|---|---|---|---|---|---|---|---|
| A | 0 | 0 | 0 | 15 | 17 | 20 | 21 | 21 | ( | 0.2 | ) |
| B | 35 | 73 | 95 | 159 | 259 | 424 | 561 | 591 | ( | 5.4 | ) |
| C | 3 | 12 | 21 | 35 | 43 | 56 | 84 | 86 | ( | 0.8 | ) |
| D | 65 | 120 | 194 | 282 | 380 | 668 | 825 | 849 | ( | 7.8 | ) |
| E | 1 | 28 | 29 | 34 | 40 | 45 | 62 | 62 | ( | 0.6 | ) |
| F | 0 | 1 | 1 | 5 | 9 | 10 | 11 | 11 | ( | 0.1 | ) |
| G | 6 | 63 | 72 | 104 | 151 | 221 | 380 | 393 | ( | 3.6 | ) |
| H | 106 | 213 | 263 | 345 | 458 | 641 | 887 | 952 | ( | 8.8 | ) |
| J | 76 | 162 | 181 | 248 | 368 | 433 | 525 | 574 | ( | 5.3 | ) |
| K | 41 | 61 | 89 | 138 | 167 | 229 | 293 | 297 | ( | 2.7 | ) |
| L | 6 | 23 | 29 | 47 | 106 | 151 | 244 | 258 | ( | 2.4 | ) |
| M | 3 | 6 | 13 | 20 | 28 | 39 | 52 | 53 | ( | 0.5 | ) |
| N | 17 | 31 | 58 | 92 | 119 | 163 | 203 | 209 | ( | 1.9 | ) |
| P | 58 | 155 | 203 | 278 | 399 | 550 | 769 | 834 | ( | 7.7 | ) |
| Q | 41 | 66 | 92 | 150 | 270 | 569 | 826 | 864 | ( | 8.0 | ) |
| R | 5 | 17 | 29 | 104 | 232 | 545 | 688 | 711 | ( | 6.5 | ) |
| S | 6 | 12 | 13 | 21 | 28 | 36 | 51 | 52 | ( | 0.5 | ) |
| T | 12 | 38 | 45 | 67 | 91 | 158 | 248 | 267 | ( | 2.5 | ) |
| U | 0 | 0 | 0 | 0 | 0 | 0 | 1 | 1 | ( | 0.0 | ) |
| Z | 4 | 15 | 23 | 30 | 32 | 47 | 56 | 57 | ( | 0.5 | ) |
| N/A | 0 | 77 | 79 | 251 | 522 | 693 | 3,523 | 3,724 | ( | 34.3 | ) |
| Total | 485 | 1,173 | 1,529 | 2,425 | 3,719 | 5,698 | 10,310 | 10,866 | ( | 100.0 | ) |

Table 3: LCC Codes

| A | General Works |
|---|---|
| B | Philosophy. Psychology. Religion |
| C | Auxiliary Sciences of History |
| D | World History and History of Europe, Asia, Africa, Australia, New Zealand, etc. |
| E | History of the Americas |
| F | History of the Americas |
| G | Geography. Anthropology. Recreation |
| H | Social Sciences |
| J | Political Science |
| K | Law |
| L | Education |
| M | Music |
| N | Fine Arts |
| P | Language and Literature |
| Q | Science |
| R | Medicine |
| S | Agriculture |
| T | Technology |
| U | Military Science |
| V | Naval Science |
| Z | Bibliography. Library Science. Information Resources (General) |

**Combination of Subjects and Languages**

The numbers and percentages of subjects for the titles written in English, French, German, Portuguese, Spanish, Italian, Dutch, and the rest ("Other") available through the DOAB as of February 24, 2018, are shown in Table 6. In Table 6, the leftmost "A" to "Z" represent the LCC subjects again. For instance, the number of titles in Social Sciences ("H") written in English and German are 606 and 132, respectively. They account for 63.5% and 13.8% of the titles in Social Sciences (the total number is 955,

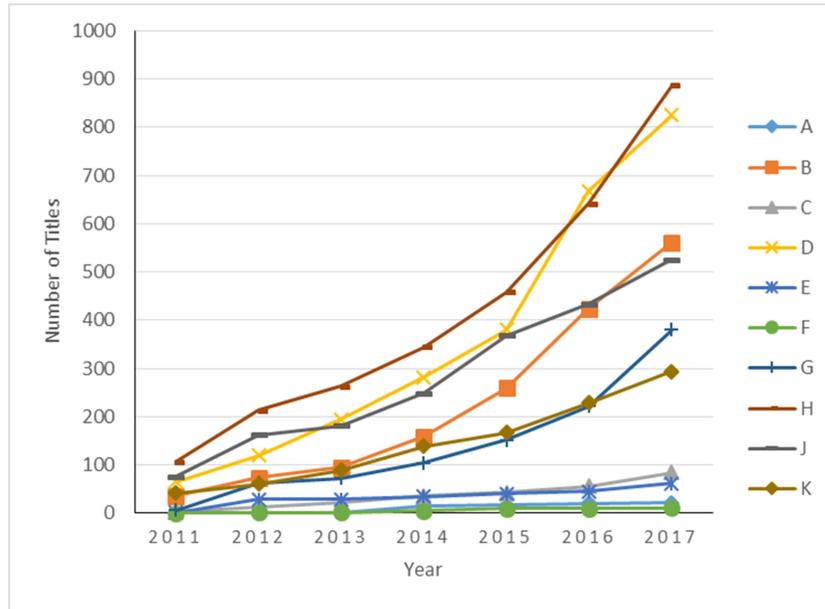

Figure 2: Changes in the Numbers of Titles of Subjects "A" to "K"

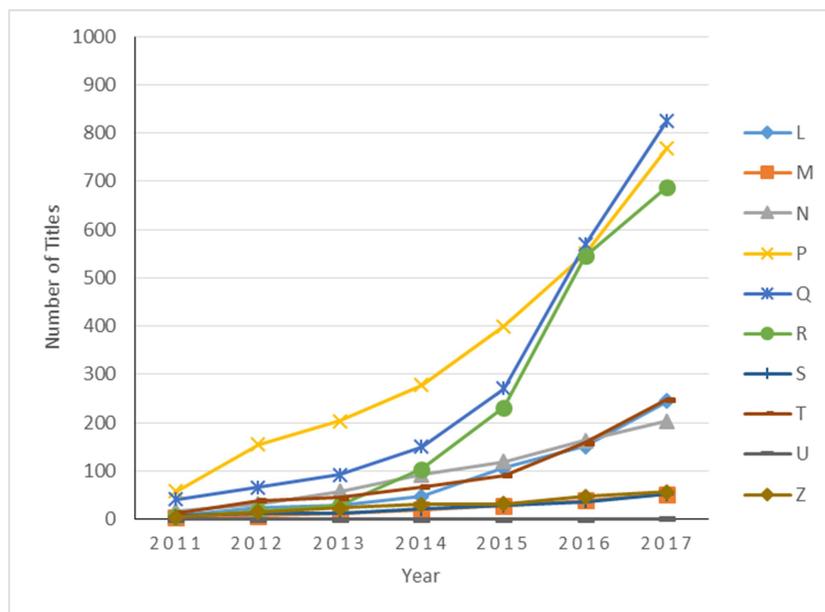

Figure 3: Changes in the Numbers of Titles of Subjects "L" to "Z"

which is shown in the rightmost column).

We can see in Table 6 that the subjects for which English is the most popular are Science ("Q"), Medicine ("R"), and Technology ("T"), which account for 85.3%, 79.2%, and 78.1%, respectively (except for minor subject "Z"). In the so-called STM (Science, Technology and Medicine) fields, English seems to be popular. The subject in which German is the most popular is Law ("K"), which accounts for 41.5%. German is relatively popular in the Fine Arts ("N") where it accounts for 31.8%, while English accounts for 41.4%. Unfortunately, many of the French books in the DOAB are not assigned LCC codes. More specifically, 2,526 French books do not have them. If they had been so assigned, we would have been able to analyze the tendency of OA French books' subjects more precisely.

**Combination of Publication and Added-on Years**

The year-wise publication and added-on numbers of titles are shown in Table 7. In Table 7, the leftmost column represents the publication years, and the top row represents the added-on years. For instance, the number of titles that were published during the period from 2005 to

Table 4: Numbers and Percentages of Titles in Each Language

| English | 5,416 | ( 49.0 ) | Afrikaans | 5 | ( 0.0 ) |
|---|---|---|---|---|---|
| French | 2,665 | ( 24.1 ) | Arabic | 5 | ( 0.0 ) |
| German | 1,062 | ( 9.6 ) | Chinese | 5 | ( 0.0 ) |
| Portuguese | 629 | ( 5.7 ) | Greek | 5 | ( 0.0 ) |
| Spanish | 429 | ( 3.9 ) | Lithuanian | 5 | ( 0.0 ) |
| Italian | 245 | ( 2.2 ) | Albanian | 4 | ( 0.0 ) |
| Dutch | 180 | ( 1.6 ) | Czech | 4 | ( 0.0 ) |
| Finnish | 31 | ( 0.3 ) | Bulgarian | 2 | ( 0.0 ) |
| Norwegian | 17 | ( 0.2 ) | Church Slavic | 2 | ( 0.0 ) |
| Latin | 10 | ( 0.1 ) | Cree | 2 | ( 0.0 ) |
| Slovene | 10 | ( 0.1 ) | Catalan | 1 | ( 0.0 ) |
| Swedish | 8 | ( 0.1 ) | Michif | 1 | ( 0.0 ) |
| Turkish | 8 | ( 0.1 ) | Ukrainian | 1 | ( 0.0 ) |
| Romansh | 7 | ( 0.1 ) | Welsh | 1 | ( 0.0 ) |
| Old Nubian | 6 | ( 0.1 ) | "Other" | 2 | ( 0.0 ) |
| Russian | 6 | ( 0.1 ) | N/A | 284 | ( 2.6 ) |
| Total | | | | 11,058 | ( 100.0 ) |

Table 5: Changes in the Numbers of Titles in Each Language

| | English | French | German | Portuguese | Spanish | Italian | Dutch | Other | N/A |
|---|---|---|---|---|---|---|---|---|---|
| 2011 | 277 | 1 | 92 | 0 | 0 | 74 | 42 | 1 | 0 |
| 2012 | 783 | 9 | 152 | 0 | 0 | 75 | 58 | 8 | 99 |
| 2013 | 970 | 13 | 303 | 0 | 2 | 87 | 64 | 10 | 2 |
| 2014 | 1,412 | 158 | 393 | 123 | 71 | 106 | 72 | 19 | 12 |
| 2015 | 2,056 | 351 | 488 | 413 | 117 | 127 | 72 | 47 | 14 |
| 2016 | 3,414 | 489 | 753 | 529 | 163 | 150 | 121 | 70 | 22 |
| 2017 | 5,086 | 2,546 | 1,014 | 605 | 427 | 234 | 179 | 129 | 125 |
| Feb. 2018 | 5,416 | 2,665 | 1,062 | 629 | 429 | 245 | 180 | 148 | 284 |

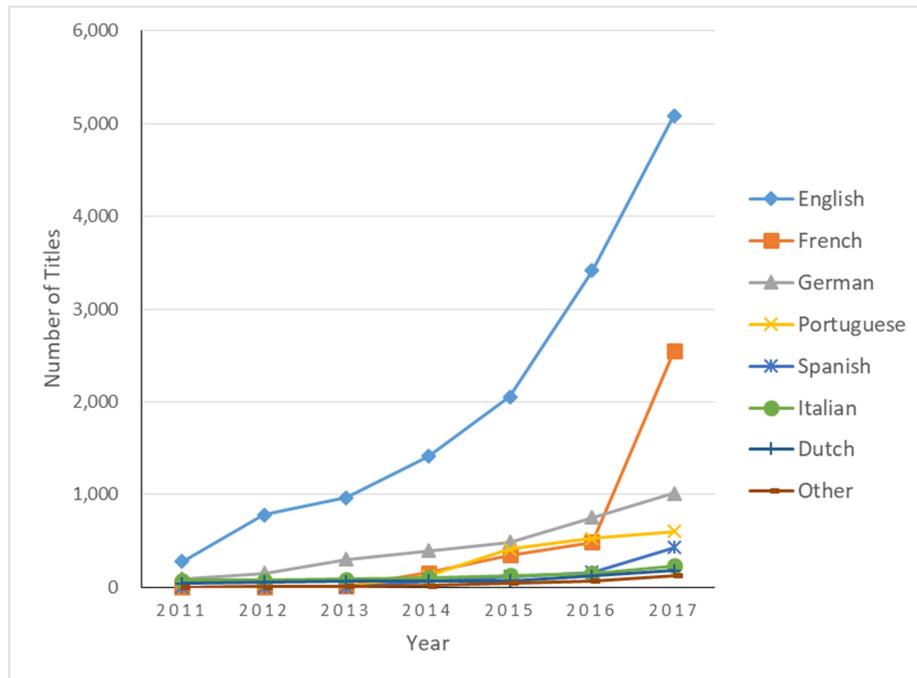

Figure 4: Changes in the Numbers of Titles in Each Language

Table 6: Numbers and Percentages of Subjects and Languages of the Titles

|   | English | French | German | Portuguese | Spanish | Italian | Dutch | Other | N/A | Total |
|---|---|---|---|---|---|---|---|---|---|---|
| A | 4 ( 18.2 ) | 1 ( 4.5 ) | 1 ( 4.5 ) | 0 ( 0.0 ) | 15 ( 68.2 ) | 1 ( 4.5 ) | 0 ( 0.0 ) | 0 ( 0.0 ) | 0 ( 0.0 ) | 22 |
| B | 370 ( 60.7 ) | 5 ( 0.8 ) | 99 ( 16.2 ) | 78 ( 12.8 ) | 14 ( 2.3 ) | 11 ( 1.8 ) | 4 ( 0.7 ) | 18 ( 3.0 ) | 11 ( 1.8 ) | 610 |
| C | 56 ( 56.0 ) | 5 ( 5.0 ) | 32 ( 32.0 ) | 0 ( 0.0 ) | 4 ( 4.0 ) | 3 ( 3.0 ) | 0 ( 0.0 ) | 0 ( 0.0 ) | 0 ( 0.0 ) | 100 |
| D | 449 ( 50.2 ) | 41 ( 4.6 ) | 242 ( 27.1 ) | 12 ( 1.3 ) | 43 ( 4.8 ) | 36 ( 4.0 ) | 49 ( 5.5 ) | 13 ( 1.5 ) | 9 ( 1.0 ) | 894 |
| E | 21 ( 33.3 ) | 0 ( 0.0 ) | 1 ( 1.6 ) | 3 ( 4.8 ) | 3 ( 4.8 ) | 1 ( 1.6 ) | 0 ( 0.0 ) | 0 ( 0.0 ) | 34 ( 54.0 ) | 63 |
| F | 0 ( 0.0 ) | 0 ( 0.0 ) | 0 ( 0.0 ) | 2 ( 18.2 ) | 8 ( 72.7 ) | 0 ( 0.0 ) | 0 ( 0.0 ) | 0 ( 0.0 ) | 1 ( 9.1 ) | 11 |
| G | 249 ( 62.7 ) | 8 ( 2.0 ) | 30 ( 7.6 ) | 5 ( 1.3 ) | 1 ( 0.3 ) | 4 ( 1.0 ) | 0 ( 0.0 ) | 4 ( 1.0 ) | 96 ( 24.2 ) | 397 |
| H | 606 ( 63.5 ) | 18 ( 1.9 ) | 132 ( 13.8 ) | 86 ( 9.0 ) | 24 ( 2.5 ) | 19 ( 2.0 ) | 31 ( 3.2 ) | 22 ( 2.3 ) | 17 ( 1.8 ) | 955 |
| J | 432 ( 75.0 ) | 6 ( 1.0 ) | 26 ( 4.5 ) | 67 ( 11.6 ) | 7 ( 1.2 ) | 12 ( 2.1 ) | 14 ( 2.4 ) | 8 ( 1.4 ) | 4 ( 0.7 ) | 576 |
| K | 91 ( 29.7 ) | 2 ( 0.7 ) | 127 ( 41.5 ) | 11 ( 3.6 ) | 49 ( 16.0 ) | 15 ( 4.9 ) | 4 ( 1.3 ) | 4 ( 1.3 ) | 3 ( 1.0 ) | 306 |
| L | 119 ( 45.9 ) | 3 ( 1.2 ) | 21 ( 8.1 ) | 62 ( 23.9 ) | 15 ( 5.8 ) | 17 ( 6.6 ) | 2 ( 0.8 ) | 11 ( 4.2 ) | 9 ( 3.5 ) | 259 |
| M | 23 ( 42.6 ) | 0 ( 0.0 ) | 16 ( 29.6 ) | 10 ( 18.5 ) | 0 ( 0.0 ) | 0 ( 0.0 ) | 2 ( 3.7 ) | 1 ( 1.9 ) | 2 ( 3.7 ) | 54 |
| N | 91 ( 41.4 ) | 2 ( 0.9 ) | 70 ( 31.8 ) | 16 ( 7.3 ) | 1 ( 0.5 ) | 9 ( 4.1 ) | 23 ( 10.5 ) | 5 ( 2.3 ) | 3 ( 1.4 ) | 220 |
| P | 526 ( 60.2 ) | 21 ( 2.4 ) | 115 ( 13.2 ) | 80 ( 9.2 ) | 11 ( 1.3 ) | 43 ( 4.9 ) | 4 ( 0.5 ) | 23 ( 2.6 ) | 51 ( 5.8 ) | 874 |
| Q | 743 ( 85.3 ) | 11 ( 1.3 ) | 50 ( 5.7 ) | 39 ( 4.5 ) | 0 ( 0.0 ) | 11 ( 1.3 ) | 2 ( 0.2 ) | 7 ( 0.8 ) | 8 ( 0.9 ) | 871 |
| R | 566 ( 79.2 ) | 5 ( 0.7 ) | 24 ( 3.4 ) | 107 ( 15.0 ) | 1 ( 0.1 ) | 2 ( 0.3 ) | 0 ( 0.0 ) | 6 ( 0.8 ) | 4 ( 0.6 ) | 715 |
| S | 38 ( 73.1 ) | 0 ( 0.0 ) | 8 ( 15.4 ) | 3 ( 5.8 ) | 0 ( 0.0 ) | 2 ( 3.8 ) | 0 ( 0.0 ) | 0 ( 0.0 ) | 1 ( 1.9 ) | 52 |
| T | 211 ( 78.1 ) | 8 ( 3.0 ) | 22 ( 8.1 ) | 17 ( 6.3 ) | 2 ( 0.7 ) | 5 ( 1.9 ) | 2 ( 0.7 ) | 0 ( 0.0 ) | 3 ( 1.1 ) | 270 |
| U | 1 ( 100.0 ) | 0 ( 0.0 ) | 0 ( 0.0 ) | 0 ( 0.0 ) | 0 ( 0.0 ) | 0 ( 0.0 ) | 0 ( 0.0 ) | 0 ( 0.0 ) | 0 ( 0.0 ) | 1 |
| Z | 40 ( 64.5 ) | 3 ( 4.8 ) | 5 ( 8.1 ) | 2 ( 3.2 ) | 2 ( 3.2 ) | 7 ( 11.3 ) | 0 ( 0.0 ) | 1 ( 1.6 ) | 2 ( 3.2 ) | 62 |
| N/A | 780 ( 20.8 ) | 2,526 ( 67.4 ) | 41 ( 1.1 ) | 29 ( 0.8 ) | 229 ( 6.1 ) | 47 ( 1.3 ) | 43 ( 1.1 ) | 25 ( 0.7 ) | 26 ( 0.7 ) | 3,746 |

Table 7: Year-wise Publication and Added-on Numbers for Titles

|  | 2011 | 2012 | 2013 | 2014 | 2015 | 2016 | 2017 | (2018) |
|---|---|---|---|---|---|---|---|---|
| ～1999 | 5 | 23 | 5 | 82 | 76 | 129 | 829 | 43 |
| 2000～04 | 135 | 34 | 19 | 74 | 76 | 57 | 530 | 17 |
| 2005～09 | 264 | 254 | 50 | 113 | 228 | 123 | 737 | 37 |
| 2010～14 | 81 | 377 | 282 | 622 | 449 | 299 | 813 | 137 |
| 2015～17 | 0 | 0 | 0 | 5 | 465 | 1,371 | 1,697 | 227 |
| (2018) | 0 | 0 | 0 | 0 | 0 | 0 | 6 | 95 |

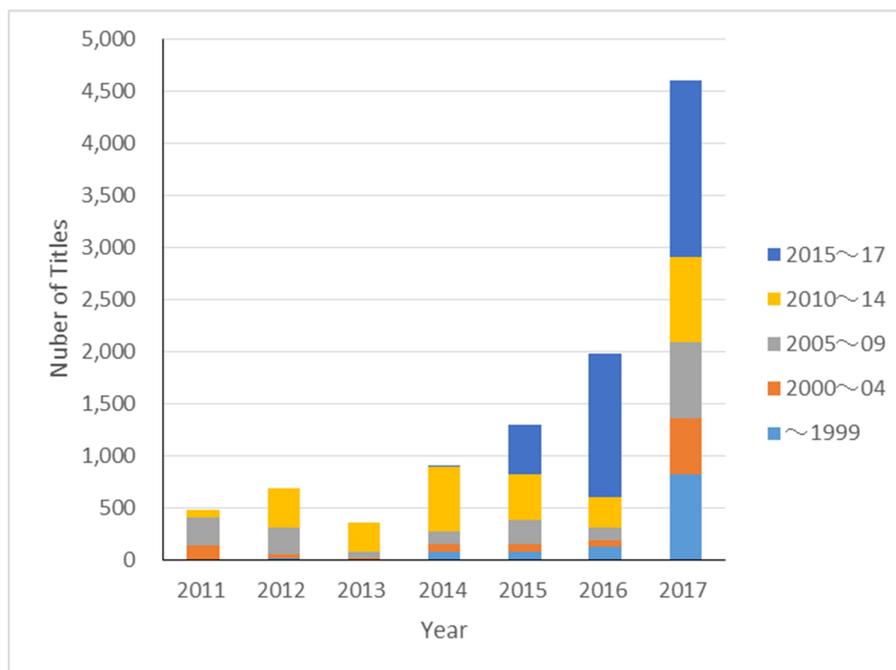

Figure 5: Numbers of Titles of Each Publication Year Added to the DOAB

2009 and that were added to the DOAB in 2012 is 254. Because our data is as of February 24, 2018, and the number of titles for 2018 is tentative (in the sense that 2018 has not ended), the numbers in the "(2018)" row and column are generally smaller than those in the other rows and columns are.

Table 8: Numbers of Titles Added to the DOAB by Each Publisher

| Publisher | 2011 | 2012 | 2013 | 2014 | 2015 | 2016 | 2017 | (2018) | Total |
|---|---|---|---|---|---|---|---|---|---|
| Frontiers Media SA | 0 | 0 | 1 | 0 | 176 | 542 | 208 | 0 | 927 |
| Presses universitaires de Rennes | 0 | 0 | 0 | 0 | 0 | 0 | 528 | 0 | 528 |
| ANU Press | 1 | 189 | 0 | 145 | 42 | 56 | 43 | 10 | 486 |
| De Gruyter | 0 | 17 | 11 | 33 | 40 | 176 | 108 | 52 | 437 |
| Springer | 0 | 10 | 17 | 26 | 25 | 64 | 207 | 1 | 350 |
| MDPI AG – Multidisciplinary Digital Publishing Institute | 0 | 0 | 0 | 2 | 50 | 92 | 134 | 53 | 331 |
| Amsterdam University Press | 136 | 42 | 26 | 9 | 13 | 33 | 46 | 9 | 314 |
| Universitätsverlag Göttingen | 101 | 40 | 39 | 25 | 1 | 57 | 30 | 3 | 296 |
| Brill | 24 | 21 | 1 | 4 | 0 | 136 | 51 | 0 | 237 |
| Böhlau | 0 | 0 | 86 | 49 | 34 | 18 | 34 | 1 | 222 |
| punctum books | 0 | 0 | 0 | 63 | 42 | 22 | 62 | 7 | 196 |
| SciELO Books – Editora UNESP | 0 | 0 | 0 | 2 | 95 | 63 | 11 | 4 | 175 |
| Bloomsbury Academic | 0 | 38 | 20 | 25 | 45 | 10 | 8 | 14 | 160 |
| Graduate Institute Publications | 0 | 0 | 0 | 0 | 0 | 0 | 147 | 0 | 147 |
| Manchester University Press | 90 | 1 | 0 | 2 | 0 | 12 | 34 | 8 | 147 |
| Other | 133 | 330 | 155 | 511 | 731 | 698 | 2,961 | 394 | 5,913 |
| Total | 485 | 688 | 356 | 896 | 1,294 | 1,979 | 4,612 | 556 | 10,866 |

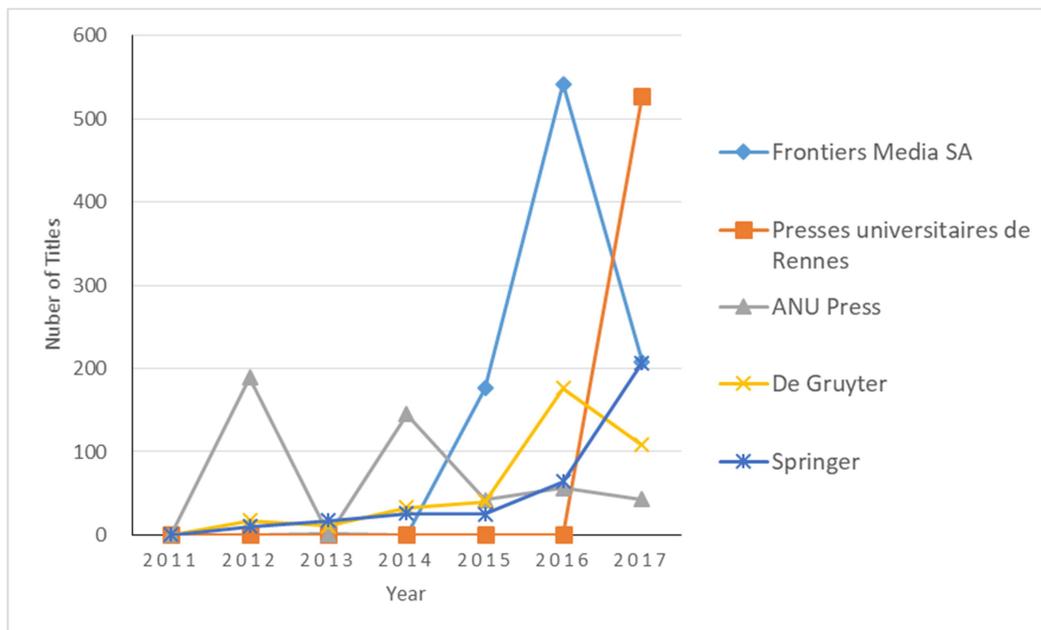

Figure 6: Numbers of Titles Added to the DOAB by the Top Five Publishers

We can see in Table 7 that some titles were added to the DOAB as OA books before being published. For instance, five titles that were published during the period from 2015 to 2017 had been added to the DOAB in 2014. It was found that 29 titles in total were published after they were added to the DOAB.[7]

The bar chart in Figure 5 shows the numbers of titles in Table 7. We can see in Figure 5 that while the ratios of titles that were added to the DOAB right after being published are high (for instance, the blue part, which represents the number of titles that were published during the period from 2015 to 2017, is the largest part of the bar representing those works that were added to the DOAB in 2017), the ratio of titles that were published in 1999 or before is on the increase, especially in 2017.

---

[7] To be precise, 29 titles have publication years that came after (and so are larger than) the added-on years. The titles that were published after being added to the DOAB in the same year are not included in these 29 titles.

Table 9: Numbers and Percentages of Subjects for the Titles Added by the Top Five Publishers

|     | Frontiers Media SA | Presses universitaires de Rennes | ANU Press | De Gruyter | Springer |
|-----|-------------------|----------------------------------|-----------|------------|----------|
| A   | 0 ( 0.0 )         | 0 ( 0.0 )                        | 0 ( 0.0 ) | 0 ( 0.0 )  | 0 ( 0.0 ) |
| B   | 92 ( 9.9 )        | 0 ( 0.0 )                        | 19 ( 3.9 )| 82 ( 18.8 )| 9 ( 2.6 ) |
| C   | 0 ( 0.0 )         | 0 ( 0.0 )                        | 16 ( 3.3 )| 9 ( 2.1 )  | 0 ( 0.0 ) |
| D   | 0 ( 0.0 )         | 0 ( 0.0 )                        | 68 ( 14.0 )| 117 ( 26.8 )| 1 ( 0.3 ) |
| E   | 0 ( 0.0 )         | 0 ( 0.0 )                        | 1 ( 0.2 ) | 1 ( 0.2 )  | 0 ( 0.0 ) |
| F   | 0 ( 0.0 )         | 0 ( 0.0 )                        | 0 ( 0.0 ) | 0 ( 0.0 )  | 0 ( 0.0 ) |
| G   | 11 ( 1.2 )        | 0 ( 0.0 )                        | 45 ( 9.3 )| 7 ( 1.6 )  | 41 ( 11.7 ) |
| H   | 0 ( 0.0 )         | 0 ( 0.0 )                        | 94 ( 19.3 )| 16 ( 3.7 )| 64 ( 18.3 ) |
| J   | 0 ( 0.0 )         | 0 ( 0.0 )                        | 100 ( 20.6 )| 4 ( 0.9 )| 9 ( 2.6 ) |
| K   | 0 ( 0.0 )         | 0 ( 0.0 )                        | 21 ( 4.3 )| 25 ( 5.7 ) | 11 ( 3.1 ) |
| L   | 0 ( 0.0 )         | 0 ( 0.0 )                        | 6 ( 1.2 ) | 4 ( 0.9 )  | 47 ( 13.4 ) |
| M   | 0 ( 0.0 )         | 0 ( 0.0 )                        | 3 ( 0.6 ) | 1 ( 0.2 )  | 0 ( 0.0 ) |
| N   | 0 ( 0.0 )         | 0 ( 0.0 )                        | 8 ( 1.6 ) | 9 ( 2.1 )  | 0 ( 0.0 ) |
| P   | 0 ( 0.0 )         | 0 ( 0.0 )                        | 23 ( 4.7 )| 56 ( 12.8 )| 4 ( 1.1 ) |
| Q   | 392 ( 42.3 )      | 0 ( 0.0 )                        | 21 ( 4.3 )| 35 ( 8.0 ) | 75 ( 21.4 ) |
| R   | 406 ( 43.8 )      | 0 ( 0.0 )                        | 39 ( 8.0 )| 15 ( 3.4 ) | 35 ( 10.0 ) |
| S   | 0 ( 0.0 )         | 0 ( 0.0 )                        | 10 ( 2.1 )| 2 ( 0.5 )  | 6 ( 1.7 ) |
| T   | 26 ( 2.8 )        | 0 ( 0.0 )                        | 7 ( 1.4 ) | 15 ( 3.4 ) | 48 ( 13.7 ) |
| U   | 0 ( 0.0 )         | 0 ( 0.0 )                        | 0 ( 0.0 ) | 0 ( 0.0 )  | 0 ( 0.0 ) |
| Z   | 0 ( 0.0 )         | 0 ( 0.0 )                        | 4 ( 0.8 ) | 21 ( 4.8 ) | 0 ( 0.0 ) |
| N/A | 0 ( 0.0 )         | 528 ( 100.0 )                    | 1 ( 0.2 ) | 18 ( 4.1 ) | 0 ( 0.0 ) |
| Total | 927 ( 100.0 )   | 528 ( 100.0 )                    | 486 ( 100.0 )| 437 ( 100.0 )| 350 ( 100.0 ) |

Table 10: Changes in the Percentages of Licensing Patterns

|            | 2011 | 2012 | 2013 | 2014 | 2015 | 2016 | 2017 | (2018) | Total |
|------------|------|------|------|------|------|------|------|--------|-------|
| CC by      | 0.1  | 0.4  | 1.1  | 1.6  | 17.8 | 41.3 | 33.7 | 4.1    | 1,682 |
| CC by-nc   | 14.4 | 4.0  | 2.8  | 7.6  | 10.2 | 27.6 | 22.6 | 10.9   | 935   |
| CC by-nc-nd| 8.6  | 12.3 | 7.7  | 8.4  | 7.4  | 16.4 | 32.7 | 6.5    | 3,139 |
| CC by-nc-sa| 0.0  | 0.0  | 2.9  | 20.9 | 36.4 | 7.2  | 30.4 | 2.1    | 994   |
| CC by-nd   | 44.8 | 20.1 | 16.7 | 5.2  | 2.3  | 6.3  | 4.6  | 0.0    | 174   |
| CC by-sa   | 0.0  | 3.5  | 3.5  | 11.9 | 4.4  | 34.4 | 28.2 | 14.1   | 227   |
| N/A        | 0.0  | 5.8  | 0.1  | 7.8  | 7.8  | 9.4  | 65.5 | 3.5    | 3,715 |
| Total      | 4.5  | 6.3  | 3.3  | 8.2  | 11.9 | 18.2 | 42.4 | 5.1    | 10,866 |

**Distribution of Publishers**

The publisher-wise numbers of titles added to the DOAB during the period from 2011 to 2017 (and from January 1 to February 24, 2018) are shown in Table 8. The leftmost column shows the publishers' names, sorted in descending order of the total number of titles shown in the rightmost column. We can see in Table 8 that the five publishers that added the most works to the DOAB are Frontiers Media SA, Presses universitaires de Rennes, ANU Press, De Gruyter, and Springer (henceforth, the "top five publishers"). While the total number of titles available through the DOAB increased at an accelerating rate (see Figure 1), the number of titles added to the DOAB by each publisher did not follow the same pattern. For instance, Presses universitaires de Rennes added all of its 528 titles in 2017, and ANU Press added 189 titles in 2012 and zero in 2013.

The numbers of titles added to the DOAB by the above-mentioned five publishers during the period from 2011 to 2017 are shown in Figure 6. We can see in Figure 6 that the numbers of titles added by Frontiers Media SA and De Gruyter decreased sharply in 2017. The number of titles added by ANU Press decreased a little. On the other hand, the number of titles added by Springer has increased constantly.

The numbers and percentages of subjects for the titles added by the top five publishers are shown in Table 9. We can see in Table 9 that more than 80% of the subjects for the titles added by Frontiers Media SA were "R" and "Q," i.e., Medicine and Science (43.8% and 42.3%, respectively). The top three subjects for the titles added by ANU Press were "J," "H," and "D," i.e., Political Science, the Social Sciences, and World History and History of Europe, Asia, Africa, Australia, New Zealand, etc. (20.6%, 19.3%, and 14.0%, respectively). As for De Gruyter, they were "D," "B," and "P," i.e., World History and History of Europe, Asia, Africa, Australia, New

Table 11: Percentages of Licensing Patterns of the Top 15 Publishers

| | CC by | CC by-nc | CC by-nc-nd | CC by-nc-sa | CC by-nd | CC by-sa | N/A | Total |
|---|---|---|---|---|---|---|---|---|
| Frontiers Media SA | 100.0 | 0.0 | 0.0 | 0.0 | 0.0 | 0.0 | 0.0 | 927 |
| Presses universitaires de Rennes | 0.0 | 0.0 | 0.0 | 0.0 | 0.0 | 0.0 | 100.0 | 528 |
| ANU Press | 0.0 | 0.0 | 0.0 | 0.2 | 0.0 | 0.0 | 99.8 | 486 |
| De Gruyter | 0.7 | 0.0 | 98.9 | 0.2 | 0.0 | 0.2 | 0.0 | 437 |
| Springer | 36.0 | 63.4 | 0.3 | 0.3 | 0.0 | 0.0 | 0.0 | 350 |
| MDPI AG – Multidisciplinary Digital Publishing Institute | 26.9 | 0.0 | 73.1 | 0.0 | 0.0 | 0.0 | 0.0 | 331 |
| Amsterdam University Press | 0.6 | 53.5 | 45.5 | 0.0 | 0.3 | 0.0 | 0.0 | 314 |
| Universitätsverlag Göttingen | 0.7 | 0.0 | 11.1 | 0.0 | 51.7 | 36.5 | 0.0 | 296 |
| Brill | 0.0 | 70.5 | 19.4 | 0.0 | 0.0 | 0.0 | 10.1 | 237 |
| Böhlau | 13.5 | 18.0 | 68.5 | 0.0 | 0.0 | 0.0 | 0.0 | 222 |
| punctum books | 0.0 | 0.0 | 0.0 | 100.0 | 0.0 | 0.0 | 0.0 | 196 |
| SciELO Books – Editora UNESP | 0.0 | 2.3 | 0.0 | 57.7 | 0.0 | 0.0 | 40.0 | 175 |
| Bloomsbury Academic | 0.0 | 8.1 | 91.3 | 0.0 | 0.0 | 0.0 | 0.6 | 160 |
| Graduate Institute Publications | 0.0 | 0.0 | 100.0 | 0.0 | 0.0 | 0.0 | 0.0 | 147 |
| Manchester University Press | 0.7 | 7.5 | 91.8 | 0.0 | 0.0 | 0.0 | 0.0 | 147 |
| Other | 8.5 | 5.2 | 28.1 | 11.7 | 0.3 | 2.0 | 44.1 | 5,913 |
| Total | 15.5 | 8.6 | 28.9 | 9.1 | 1.6 | 2.1 | 34.2 | 10,866 |

Zealand, etc.; Philosophy, Psychology, Religion; and Language and Literature (26.8%, 18.8%, and 12.8%, respectively). From this result, it can be said that Frontiers Media SA, ANU Press, and De Gruyter were mainly adding STM, Social Sciences, and Humanities titles, respectively. As for Springer, the top three subjects for its titles were "Q," "H," and "T," i.e., Science, Social Sciences, and Technology (21.4%, 18.3%, and 13.7%, respectively). Springer is adding both STM and Social Sciences titles.

**Licensing Patterns**

Chronological changes in the percentages of the licensing patterns are shown in Table 10.[8] We can see in Table 10 that the most popular pattern is "CC by-nc-nd" (3,139 titles), which increases from 2015 to 2017 (i.e., from 7.4% to 32.7%). The second-most popular pattern is "CC by" (1,682 titles), which increases from 2011 to 2016 (i.e., from 0.1% to 41.3%). Note that "CC by" is the most accommodating licensing pattern, which requires only attribution display and even allows commercial re-use.

The distribution of the licensing patterns of the top 15 publishers is shown in Table 11. We can see in Table 11 that the licensing pattern preferred by each publisher differs significantly. For instance, Frontiers Media SA prefers "CC by" (100% of their 927 titles belong to this pattern). De Gruyter and Springer prefer "CC by-nc-nd" and "CC by-nc," respectively (98.9% and 63.4% of their titles belong to these patterns). Quite a few publishers adopt more than one licensing pattern. For instance, Springer is adopting "CC by-nc" (63.4%) and "CC by" (36.0%), while AUP is adopting "CC by-nc" (53.5%) and "CC by-nc-nd" (45.5%). It would be interesting to pursue the criteria that are leading the publishers to adopt different patterns. The publishers other than these top 15 publishers (in the "Other" row) prefer "CC by-nc-nd" most of all (28.1% of 5,913 titles); their numbers account for more than half of all the 3,139 titles of "CC by-nc-nd" in Table 11.

**Authors and Editors**

Below, we will first describe the number of authors or editors shown in the front pages of 40 randomly selected English titles. Then, we will describe the affiliations and positions of their first author or editor.

Table 12 shows the number of authors or editors identified in the front pages. It was found that 15 titles (37.5%) were written by just one author. They did not have editors. By contrast, 24 titles (60.0%) had one or more editors. For instance, we can see in Table 12 that eight titles (20.0%) had three editors. Such titles often had many authors.

The affiliations and positions of the first authors or editors are shown in Table 13. We were unable to find affiliations for four (10.0%): J. H. M. C. Boelaars of *Head-Hunters about Themselves*, Christa Jungnickel of

---

[8] "by," "nc," "nd," and "sa" represent forms of Creative Commons licenses: "by" (you must give appropriate credit, provide a link to the license, and indicate if changes were made; you may do so in any reasonable manner, but not in any way that suggests the licensor endorses you or your use); "Non Commercial" or "nc" (you may not use the material for commercial purposes); "No Derivatives" or "nd" (if you remix, transform, or build upon the material, you may not distribute the modified material); and "Share Alike" or "sa" (if you remix, transform, or build upon the material, you must distribute your contributions under the same license as the original), respectively. <https://creativecommons.org/licenses/>

Table 12: Numbers of Authors or Editors

|        | 1          | 2         | 3         | 4         | Total      |
|--------|------------|-----------|-----------|-----------|------------|
| Author | 15 ( 37.5 ) | 1 ( 2.5 ) | 0 ( 0.0 ) | 0 ( 0.0 ) | 16 ( 40.0 ) |
| Editor | 4 ( 10.0 )  | 9 ( 22.5 ) | 8 ( 20.0 ) | 3 ( 7.5 )  | 24 ( 60.0 ) |

Table 13: Affiliations of the First Authors or Editors

| University Faculties | Professor           | 11 ( 27.5 )  |
|----------------------|---------------------|--------------|
|                      | Associate Professor | 3 ( 7.5 )    |
|                      | Assistant Professor | 2 ( 5.0 )    |
|                      | Reader              | 3 ( 7.5 )    |
|                      | Other               | 10 ( 25.0 )  |
| Institute Researchers |                    | 5 ( 12.5 )   |
| Museum Curator        |                    | 1 ( 2.5 )    |
| Ph.D. Student         |                    | 1 ( 2.5 )    |
| None                  |                    | 4 ( 10.0 )   |
| Total                 |                    | 40 ( 100.0 ) |

*Cavendish: The Experimental Life,* Joshua Rothe of *An Unspecific Dog: Artifacts of This Late Stage in History,* and Michael Munro of *Of Learned Ignorance: Idea of a Treatise in Philosophy*. Boelaars states in the preface of his book that the first summary of it appeared in 1958, i.e., 60 years ago. He may therefore have retired and may not have any affiliation. Jungnickel is presented in the preface to her book as the wife of the second author of the book. Because the second author was a professor emeritus, she may also be old enough not to have any institutional affiliation. As for Rothe and Munro, we could not find further information (it is thus possible that they have some affiliation). From these results, we can say that 35 individuals (i.e., 40 minus the above-mentioned four and one Ph.D. student in Table 13) out of 40 (87.5%) belong to so-called *sound* institutions, such as universities, as employees and thus that they are probably receiving salaries that are sufficient to live on. Only such people can make their works OA, and it is possible that only such people can conduct research.

**Availability of Contents as Text Format**

It was found that all 40 English titles we investigated were in PDF, HTML, or EPUB format and that the contents were available in text format. Such texts increase the opportunities for re-use, in line with the purpose of OA. Various applications are possible, such as use in book recommendation systems.

**DISCUSSION**

As it can be seen in Figure 1, the number of OA books available through the DOAB is increasing at an accelerating rate. The DOAB is a representative directory of OA books. The results seem to indicate that the number of OA books in the world is increasing and such an approach to publishing is prevailing. Since many OA books are peer reviewed, and in that sense, reliable academic books, libraries should consider utilizing them. Adding the OA books to libraries' OPACs (or discovery services) and providing them in library collections are interesting directions.

As it can be seen in Table 7, some books became OA before being published. It would be interesting to determine whether publishers are making books public on the Web first, free of charge, and then deciding to publish (or not publish) as printed books depending on users' reactions. Ferwerda et al. (2013) reported that making books OA had no negative effects on their sales.

As it can be seen in Figure 5, the number and ratio of OA books that were originally published in 1999 or before are increasing. It would be interesting to explore whether these books are now out of print and publishers have decided to disseminate them again as OA books. Note that there are many valuable books that have unfortunately gone out of print (we leave it for future research to investigate whether they are actually out of print or not).

As it can be seen in Table 2, the three most popular subjects among OA books were Social Sciences (952 titles of LCC "H"), Science (864 titles of LCC "Q"), and World History and History of Europe, Asia, Africa, Australia, New Zealand, etc. (849 titles of LCC "D"). The most popular three languages in which the OA books were written were English (5,416 titles), French (2,665 titles), and German (1,062 titles) (see Table 4). However, this order may soon change because (1) the number of French titles was heavily influenced by Presses universitaires de Rennes, which abruptly made 928 French titles available in 2017 (French was the fourth most popular language in 2016; see Table 5); (2) no LCC categories were assigned to these 928 titles (i.e., they may be "H," "Q," "D," or other); and (3) Frontiers Media SA, the publisher that registered the most OA books in general[9] and as many as 392 OA books in the field of Science ("Q"), sharply reduced the number of OA books to be added in 2017 (down to 208 from 542 in 2016). If these publishers change their policies, the statistics on the languages and subjects of the DOAB might change.

**CONCLUSION**

In the present paper, OA books available through the DOAB were investigated and the number of titles, the distribution of subjects, languages, publishers, publication years, licensing patterns, etc., were clarified. Their chronological changes were also shown. The results showed that OA books were increasing at an accelerating rate. As for the distribution of subjects, Social Sciences ("H" in LCC), Science ("Q"), and World History and History of Europe, Asia, Africa, Australia, New Zealand, etc. ("D") were the most popular. With regard to languages, English, French, and German were the most common. As for publishers, Frontiers Media SA, Presses universitaires de Rennes, and ANU Press had registered more books in the DOAB than any other publishers. Many books were newly published, but older books that had been published in 1999 or before have also begun to be available recently. As for the licensing patterns, "CC by-nc-nd" and "CC by" were the most popular. Considering these tendencies, libraries should begin to utilize OA books by, for instance, providing them as part of their collections.

As future research, we would like to investigate (1) the distribution of the levels of OA books (e.g., identifying how many books are suitable as textbooks for undergraduate students, how many are for expert researchers, and in which subjects) and (2) the business models adopted for each OA book (e.g., identifying who paid to publish it). Through these studies, better ways of utilizing and producing OA books will be clarified.

**REFERENCES**

[1]　Adema, J. (2010) *Overview of Open Access Models for eBooks in the Humanities and Social Sciences*, OAPEN, 73p. Retrieved 6 March 2018 from: <https://curve.coventry.ac.uk/open/file/a976330e-ed7a-4bd5-b0ed-47cab90e9a5e/1/ademaoapen2comb.pdf>

[2]　Adema, J. (2012) *DOAB User Needs Analysis: Final Report*, 77p. Retrieved 6 March 2018 from: <https://doabooks.files.wordpress.com/2012/11/doab-user-needs.pdf>

[3]　Amano, E. (2017) "Trends in Open Access Monographs in Europe, Current Awareness," (333), p. 12-16. Retrieved 6 March 2018 from: <https://doi.org/10.11501/10955543> (Text in Japanese)

[4]　Barnes, C. et al. (2017) *Surveying the Scalability of Open Access Monograph Initiatives: Final Report*, Michigan Publishing, 90p. Retrieved 6 March 2018 from: <https://deepblue.lib.umich.edu/handle/2027.42/139888>

[5]　Chakrabarti, A. and Mandal, S. (2017) "Overview of Open Access Books in Library and Information Science in DOAB," International Journal of Library and Information Studies, 7(4), p. 185–192. Retrieved 9 May 2018 from: <http://ijlis.org/img/2017_Vol_7_Issue_4/185-192.pdf>

[6]　Collins, E. and Milloy, C. (2012) "A Snapshot of Attitudes towards Open Access Monograph Publishing in the Humanities and Social Sciences: Part of the OAPEN-UK Project," Insights: The UKSG Journal, 25(2), p.192-197. Retrieved 15 May 2018 from: <https://doi.org/10.1629/2048-7754.25.2.192>

[7]　Collins, E. and Milloy, C. (2016) *OAPEN-UK Final Report: A Five-year Study into Open Access Monograph Publishing in the Humanities and Social Sciences*, Arts & Humanities Research Council, 94p. Retrieved 6 March 2018 from: <http://oapen-uk.jiscebooks.org/files/2016/01/OAPEN-UK-final-report.pdf>

[8]　Crossick, G. (2015) *Monographs and Open Access: A Report to HEFCE*, HEFCE, 77p. Retrieved 6 March 2018 from: <http://www.hefce.ac.uk/media/hefce/content/pubs/indirreports/2015/Monographs,and,open,access/2014_monographs.pdf>

[9]　Ferwerda, E. (2014) "Open Access Monograph Business Models," Insights: The UKSG Journal, 27(S), p. 35-38. Retrieved 6 March 2018 from: <http://doi.org/10.1629/2048-7754.46>

[10]　Ferwerda, E. et al. (2013) *OAPEN-NL: A Project Exploring Open Access Monograph Publishing in the Netherlands: Final Report*. OAPEN Foundation, 96p. Retrieved 6 March 2018 from: <https://www.researchgate.net/publication/273450141_OAPEN-NL_-_A_project_exploring_Open_Access_monograph_publishing_in_the_Netherlands_Final_Report>

[11]　Hacker, A. (2014) "Building it Together: Collaboration in University-based Open Access Book Publishing," Insights: The UKSG Journal, 27(S), p. 26–29. Retrieved 6 March 2018 from: <http://dx.doi.org/10.1629/2048-7754.120>

[12]　Hacker, A. and Corrao, E. (2017) "Laying Tracks as the Train Approaches: Innovative Open Access Book Publishing at Heidelberg University from the Editors' Point of View," Journal of Scholarly Publishing, 48(2), p. 76–89. Retrieved 6 March 2018 from: <https://doi.org/10.3138/jsp.48.2.76>

[13]　IFLA (2011) *IFLA Statement on Open Access: Clarifying IFLA's Position and Strategy*. Retrieved 6

---

[9] Khanchandani and Kumar (2017) state that it was the most active OA book publisher.

March 2018 from: <https://www.ifla.org/files/assets/hq/news/documents/ifla-statement-on-open-access.pdf>

[14] Jisc Collections and OAPEN Foundation (2016) *Investigating OA Monograph Services: Final Report*, 13p. Retrieved 6 March 2018 from: <https://www.oapen.org/content/sites/default/files/u6/Jisc-OAPEN%20pilot%20Final%20report.pdf>

[15] Karak, S. and Mandal, S. (2017) "Library and Information Science E-Books through Directory of Open Access Books (DOAB): A Case Study," International Research: Journal of Library & Information Science, 7(4), p. 729-734. Retrieved 9 May 2018 from: <http://irjlis.com/wp-content/uploads/2018/01/11-IR435-74.pdf>

[16] Khanchandani, V. and Kumar, M. (2017) Mapping of E-books in Science & Technology: An Analytical Study of Directory of Open Access Books. DESIDOC Journal of Library and Information Technology, 37(3), p. 172–179. Retrieved 9 May 2018 from: <http://publications.drdo.gov.in/ojs/index.php/djlit/article/view/10692>

[17] Lamani, M. B. et al. (2018) "Open Access E-books in Science and Technology: A Case Study of Directory of Open Access Books," DESIDOC Journal of Library and Information Technology, 38(2), p. 141-144. Retrieved 9 May 2018 from: <http://publications.drdo.gov.in/ojs/index.php/djlit/article/view/8494>

[18] London Economics (2015) *Economic Analysis of Business Models for Open Access Monographs: Annex 4 to the Report of the HEFCE Monographs and Open Access Project*, Higher Education Funding Council for England, 38p. Retrieved 6 March 2018 from: <http://www.hefce.ac.uk/media/hefce/content/pubs/indirreports/2015/Monographs,and,open,access/2014_monographs4.pdf>

[19] Mongeau, P. (2018) "'The Future is Open?': An Overview of Open Access Monograph Publishing," The iJournal, 3(2), p. 1-10. Retrieved 9 May 2018 from: <http://www.theijournal.ca/index.php/ijournal/article/view/29481/21972>

[20] Montgomery, L. et al. (2017) *Exploring the Uses of Open Access Books via the JSTOR Platform*, KU Research Press, 43p. Retrieved 6 March 2018 from: http://dx.doi.org/10.17613/M6CV52

[21] Nagaraja, A. and Clauson, K. A. (2009) "Database Coverage and Impact Factor of Open Access Journals in Pharmacy," Journal of Electronic Resources in Medical Libraries, 6(2), p. 138-145. Retrieved 6 March 2018 from: <https://doi.org/10.1080/15424060902932219>

[22] Neylon, C. et al. (2018) *The Visibility of Open Access Monographs in a European Context European Context: A Report Prepared by Knowledge Unlatched Research*, Knowledge Unlatched Research, 44p. Retrieved 9 May 2018 from: <http://dx.doi.org/10.17613/M6156F>

[23] Shore, E. (2017) AAU, ARL, AAUP to Launch Open Access Monograph Publishing Initiative. Retrieved 6 March 2018 from: <http://www.arl.org/news/arl-news/4243-aau-arl-aaup-to-launch-open-access-monograph-publishing-initiative-project-will-share-scholarship-freely-more-broadly#.WurimojFJPb>

[24] Snijder, R. (2013) "A Higher Impact for Open Access Monographs: Disseminating through OAPEN and DOAB at AUP," Insights: The UKSG Journal, 26(1), p. 55-59. Retrieved 6 March 2018 from: <https://doi.org/10.1629/2048-7754.26.1.55>

[25] Snijder, R. (2016) "Revisiting an Open Access Monograph Experiment: Measuring Citations and Tweets 5 Years Later," Scientometrics, 109(3), p. 1855-1875. Retrieved 6 March 2018 from: <https://doi.org/10.1007/s11192-016-2160-6>

[26] Speicher, L. (2016) "Does Publishing a Book as Open Access Affect Print Sales?" TXT, 2016(1), p. 124-127. Retrieved 6 March 2018 from: <https://openaccess.leidenuniv.nl/handle/1887/42723>

[27] Speicher, L. (2017) "Open Access Monographs: Current UK University Press Landscape," Interscript Online Magazine. Retrieved 6 March 2018 from: <https://www.interscriptjournal.com/online-magazine/open-access-monographs>

[28] Tanabe, Y. (2015) "Springer's Scholarly Books and Open Access," Journal of Information Processing and Management, 57(11), p. 818-825. Retrieved 6 March 2018 from: <https://doi.org/10.1241/johokanri.57.818> (Text in Japanese)

[29] UNESCO (1964) Recommendation Concerning the International Standardization of Statistics Relating to Book Production and Periodicals. Retrieved 6 March 2018 from: <http://portal.unesco.org/en/ev.php-URL_ID=13068&URL_DO=DO_TOPIC&URL_SECTION=201.html>